\documentclass[aps,twocolumn]{revtex4}
\usepackage{epsfig}
\usepackage{graphicx}
\usepackage{amsmath,amssymb,color}
\usepackage[english]{babel}

\parskip=\medskipamount

\newcommand{\eq}[1]{(\ref{#1})}
\newcommand{\fig}[1]{Fig.~\ref{#1}}

\newcommand{\be}{\begin{equation}}
\newcommand{\ee}{\end{equation}}

\begin{document}

\title{Field-driven tracer diffusion through curved bottlenecks: Fine structure of first passage events}

\author{A. Valov,$^{a}$, V. Avetisov,$^{a}$ S. Nechaev,$^{b,c}$ and G. Oshanin,$^{d}$}

\affiliation{$^{a}$~N.N. Semenov Institute of Chemical Physics RAS, 119991 Moscow, Russia \\ $^{b}$~Interdisciplinary Scientific Center Poncelet (CNRS UMI 2615), 119002 Moscow, Russia \\ $^{c}$~P.N. Lebedev Physical Institute RAS, 119991 Moscow, Russia \\ $^{d}$~Sorbonne Universit\'e, CNRS, Laboratoire de Physique Th\'eorique de la Mati\`{e}re Condens\'ee, LPTMC (UMR CNRS 7600), 4 Place Jussieu, CEDEX 05, 75252 Paris, France}

\begin{abstract}

Using scaling arguments and extensive numerical simulations, we study dynamics of a tracer particle in a corrugated channel represented by a periodic sequence of broad chambers and narrow funnel-like bottlenecks enclosed by a hard-wall boundary. A tracer particle is affected by an external force pointing along the channel, and performs an unbiased diffusion in the perpendicular direction. We present a detailed analysis a) of the distribution function of the height above the funnel's boundary at which the first crossing of a given bottleneck takes place, and b) of the distribution function of the first passage time to such an event. Our analysis reveals several new features of the dynamical behaviour which are overlooked in the studies based on the Fick-Jacobs approach. In particular, trajectories passing through a funnel concentrate predominantly on its boundary, which makes first-crossing events very sensitive to the presence of binding sites and a microscopic roughness.

\end{abstract}

\maketitle

\section{Introduction}

Particles dynamics in channels with a spatially varying cross-section is a challenging field of research due to applications in diverse systems spread across several disciplines -- physics, chemistry, chemical physics, molecular or cellular biology (see, e.g.\cite{Burada2007, Burada2009, mal0,rubi,Rout,K}), as well as in man-designed micro-fluidic devices\cite{Stone,Granick}. Typically, such channels can be viewed as a sequence of sufficiently broad chambers with impenetrable hard walls, followed by bottlenecks in form of narrow curved funnels, which permit for a passage from one chamber to another. In turn, particles can be simple molecules\cite{Burada2007, Burada2009,mal0,rubi}, or polymers \cite{pal,sak1,bianco}, may be present in tiny amounts, or
at finite concentrations thus forming a confined crowding environment\cite{zilm,benichou0, benichou,clog,active}. The particle may move under internal stimuli only, due to interactions
with the solvent molecules and thus execute unbiased diffusion\cite{mal0,rubi}, or be manipulated externally and perform a diffusion with a drift due to an imposed electric or magnetic fields. \cite{Burada2007,Burada2009,benichou0,benichou,clog,active,mal}. Among a wide variety of relevant settings, one can highlight representative examples provided by a transport of neurotransmitters in synapses\cite{Schuss2007}, DNA sequencing in corrugated pores\cite{Chinappi2018}, and processes of a viral DNA/RNA injection into cells through a narrow pore which connects the viral capsid to the cell membrane \cite{mar}. Plenty of other examples, theoretical concepts, numerical and experimental analyses have been summarised in reviews\cite{Burada2009,mal0,rubi} and in a recent special issue\cite{mal3}.

From a theoretical side, the problem is, of course, too complex to be tackled analytically in full detail. The only available analytical descriptions rely on the so-called Fick-Jacobs (FJ) approach\cite{FJ,zwan} and its subsequent generalisations\cite{Burada2007, mal0,rubi,mal, KP2006, Mangeat2017}. The FJ approach consists in an effective reduction of the original multidimensional system to a problem of a one-dimensional diffusion in presence of a confining potential, which mimics a spatial variation of the boundaries. Such an approximation is certainly physically meaningful and permits to gain a global insight into the behaviour of important characteristic properties in realistic systems, such as, e.g.  particles currents across the channel, mean first passage times from a chamber to a chamber\cite{Burada2007,mal0,rubi,mal,KP2006, Mangeat2017}, and even to quantify  fluctuations of the first passage times\cite{mal4}. On the other hand, effects which may take place due to a diffusive motion in the direction perpendicular to the principal axis of a channel are clearly disregarded within such an approach.

In this paper, we study dynamics in a two-dimensional channel with a periodically-modulated impenetrable boundary from a perspective mentioned above, focussing on fine details of the dynamics beyond the FJ picture. We consider a typical set-up of the so-called \textit{constant force} active micro-rheology\cite{poon}, in which a tracer particle -- a charge carrier or a magnetic bead -- experiences an action of a constant force pointing along the channel. The tracer particle is subject to random thermal forces due to interactions with the host medium - an incompressible solvent, which acts as a heat bath. Hence, in our settings, the tracer particle performs an unbiased diffusion in the direction perpendicular to an applied force, and a biased diffusion along the field. Evidently, in channels with a spatially-varying cross-section the longitudinal and normal components of the tracer particle motion at any instantaneous position are effectively coupled due to interactions with a curved boundary. Using scaling arguments and extensive numerical simulations, we analyse an impact of this circumstance on the dynamical behaviour and more specifically, on fine details of first crossing events of a given funnel-like bottleneck.

Our first focal question concerns the height above the funnel's boundary at which a crossing of the most narrow part of a given bottleneck occurs for the first time. Our analysis reveals that, rather surprisingly, once the starting point is right on the principal axis of the channel and is sufficiently far from a (not very narrow) bottleneck, the first crossing of a given bottleneck  takes place not in the centre of the funnel, as one can intuitively expect, but very close to either upper or lower boundaries. The corresponding distribution of the crossing height has thus pronounced maxima at the location of both boundaries and a deep in the centre of a channel, such that a crossing at the principal axis of a channel is the least probable event.

When the starting point lies on the principle axis but is sufficiently close to the bottleneck centre, the distribution of the crossing height may have a three-modal form, with two peaks at the boundaries and a peak in the centre. For starting points located right at the boundary, the distribution is always strongly skewed -- it has a single maximum at the corresponding
boundary and vanishes with the distance from it. In other words, regardless of the precise location of a starting point, a biased tracer particle predominantly moves along the boundaries of a funnel, which is thus a generic behaviour. This effect can be certainly very important in case when boundaries of a curved funnel contain chemically-active species to which the tracer particle can bind\cite{Schuss2007} -- clearly, such a motion enhances the binding probability. Even more important, in realistic systems, the channel's boundaries are typically rough on a microscopic scale. The trajectories forced to move along the boundaries have to probe their rough landscape, which exerts a higher frictional force on the tracer particle,
than the solvent molecules do in case of the motion through the solvent. This implies that for a field-driven passage through a bottleneck not the bulk mobility is a relevant parameter, but rather the mobility for motion along the boundary.

Another characteristic property which we analyse is the first passage time to a most narrow part of a given bottleneck, regardless of the precise height at which this event happens. We construct in our work the full distribution function of such times, discuss its dependence on the starting point, and also probe its effective broadness. We specify conditions at which it suffice to know  the \textit{mean} values of such times only, (which are thus representative of an actual behaviour), and when this knowledge is insufficient, i.e. when the first passage times are very defocused and thus one needs to know a full distribution function.

The paper is structured as follows. In Section \ref{sys} we describe the model system and introduce basic notations. In Section \ref{results} we summarise our main results. Lastly, in Section \ref{conc} we conclude with a brief recapitulation of our findings and outline some perspectives for a further research.

\section{System and methods}
\label{sys}

\subsection{The model}

\begin{figure}[ht]
\epsfig{file=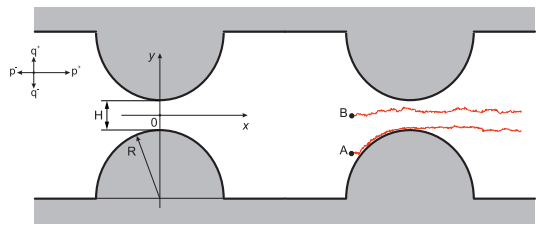,width=7cm}
\caption{A schematic view of system's geometry. Two-dimensional corrugated channel represented as a periodic sequence of chambers and bottlenecks -- funnels formed by two symmetric semicircles of radius $R$ separated  by distance $H$ from their apex points. $Ox$-axis coincides with the principal axis of a channel. A tracer particle starts from either position $A$ (placed near the funnel's boundary), or from the point $B$ situated right on the principal axis of a channel some distance $L$ apart from the most narrow part of a funnel. The tracer particle is biased by an external force to move preferentially in the positive $Ox$-direction, and performs an unbiased diffusive motion along the $Oy$-axis. The parameters $(p,q)$ on the left of the figure indicate the amplitudes of the respective  transition probabilities. Bias along the $x$-axis motion takes place once $p^+>p^-$. Red noisy curves starting from the points $A$ and $B$ show two typical  realisations of  trajectories of a tracer particle.}
\label{fig01}
\end{figure}

Consider a two-dimensional corrugated channel enclosed by impermeable hard walls, i.e. there are no other interactions with the walls apart of a hard-core exclusion. We stipulate that the $Ox$-axis coincides with the principal axis of a channel, while $Oy$-axis goes across the channel. The cross-section of a channel is a periodic function of the coordinate $x$, and such a periodic corrugation is modelled by placing, as depicted in Fig. \ref{fig01}, semicircles $W_s(x)$, where
\be
W_s(x)=\sqrt{R^2-x^2}-R-\frac{H}{2} \qquad \mbox{for $-R\leq x\leq 0$} \,,
\label{mod01}
\ee
symmetrically with respect to the $Ox$ axis. The distance between the tip points of the semicircles
(i.e. the thickness of the most narrow part of a funnel) is denoted as $H$. The maximal thickness of chambers is thus equal to $2R + H$.

We suppose next that a channel is filled with an incompressible solvent, which acts as a heat bath and maintains a constant temperature in the system. A tracer particle is introduced into a system and is placed initially either very close to the boundary of a funnel, away from the most narrow part of a bottleneck, or at the principal axis of a channel some distance $L$ apart of a bottleneck (see points $A$ and $B$ in Fig. \ref{fig01}, respectively). A tracer particle interacts with the solvent molecules and experiences a constant drift along the $Ox$ axis. Overall, it performs an unbiased diffusion along the $Oy$ axis, and a biased diffusive motion along the $Ox$ axis. In our simulations, we appropriately discretise time such that at each tick of a clock a tracer particle chooses at random a jump along $0y$ or $Ox$ axes with the corresponding jump probabilities $P_{Oy}(dy)$ and $P_{Ox}(dx)$ given by
\be
\begin{cases}
P_{Ox}(dx)\sim e^{\frac{(dx-\mu)^2}{2\sigma^2}} & \mbox{for displacement along $Ox$} \medskip \\ P_{Oy}(dy)\sim e^{\frac{dy^2}{2\sigma^2}} & \mbox{for displacement along $Oy$} \,,
\end{cases}
\label{mod02}
\ee
where $\sigma$ is a characteristic length scale and $\mu$ is a reduced force. We keep track of all individual realisation of the tracer particle trajectories. We record time moments when the tracer particle arrives for the first time to the most narrow part of a funnel, as well as the height $h$ above the funnel's boundary at which a first passage event takes place. This permits us to eventually construct the distribution function of the time of a first crossing event, as well as the distribution function of the height $h$.

\subsection{Applicability region of the FJ approach}

To set up a scene, it might be instructive to define first some characteristic time scales. Following Ref. \cite{Burada2007}, we quantify dynamics of a tracer particle in longitudinal (along the $Ox$-axis) and tangential (along the $Oy$-axis) directions, by introducing the following characteristic times:
\be
\begin{cases}
\tau_x=\frac{\Delta x^2}{2D}=\frac{L^2}{\sigma^2} & \quad \text{typical diffusion time along $Ox$} \medskip \\ \tau_y=\frac{\Delta y^2}{2D}=\frac{w(x)^2}{\sigma^2} & \quad \text{typical diffusion time along $Oy$} \medskip \\ \tau_d=\frac{L}{\mu} & \quad \text{typical drift time  along $Ox$}
\end{cases}
\label{fj01}
\ee
where $w(x)=2R+H-2\sqrt{R^2-x^2}$ (for $|x|\le R$) is  the thickness of a channel at position $x$. Here, $\tau_x$ a typical time needed for an unbiased diffusive motion to travel on distance $L$,  $\tau_y$ is a typical time needed to reach the boundary of a channel, while $\tau_d$ corresponds to a biased motion and is a time needed to move ballistically on distance $L$, under an action of a reduced force $\mu$.

The FJ approximation relies on the assumption of an equilibration of dynamics in the transverse direction. For unbiased systems, such an equilibration happens if the transverse characteristic time scale is much less than the characteristic time for longitudinal diffusion:
\be
\frac{\tau_y}{\tau_x}\sim\left(\frac{1}{2}w'(x)\right)^2\ll 1
\label{fj02}
\ee
In presence of an external force, an equilibration of dynamics in the $Oy$-direction takes place if:
\be
\frac{\tau_y}{\tau_d}\sim\left(\frac{\mu w^2(x)}{\sigma^2 L}\right)\ll 1
\label{fj03}
\ee
According to Ref.\cite{Burada2007}, one defines the "composite" parameter, $\eta$, which demarcates
the range of validity of the FJ approximation. This parameter is given by
\be
\eta=\left<\left(\frac{1}{2}w'(x)\right)^2\right>+\left<\frac{\mu w^2(x)}{\sigma^2 L} \right>
\label{fj04}
\ee
where averaging is performed along the $Ox$-axis over an interval ranging from the starting point till the target -- a centre of the most narrow part of a bottleneck. Condition  $\eta \ll 1$ defines a region in the parameter space, in which the FJ approximation is justified.

\section{Results}
\label{results}

\subsection{Scaling estimates}

\begin{figure}[ht]
\epsfig{file=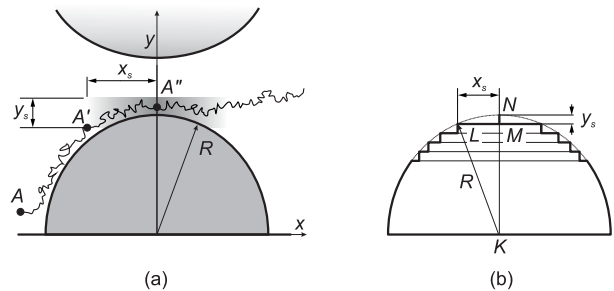,width=8cm}
\caption{(a) Two-dimensional biased random walk starting close to the funnel's boundary; (b) An auxiliary geometric construction used to derive \eq{sc02}.}
\label{fig02}
\end{figure}

We focus on a typical behaviour of the trajectories of a tracer particle which start at the point $A$ close to one of the channel's walls. In this case, evidently, the external field acting on a tracer particle will cause most strong interactions with the wall, forcing any trajectory to remain within some close vicinity (a boundary layer) of a curved boundary. Such a typical trajectory is depicted in \fig{fig02}a.

\begin{figure}[ht]
\epsfig{file=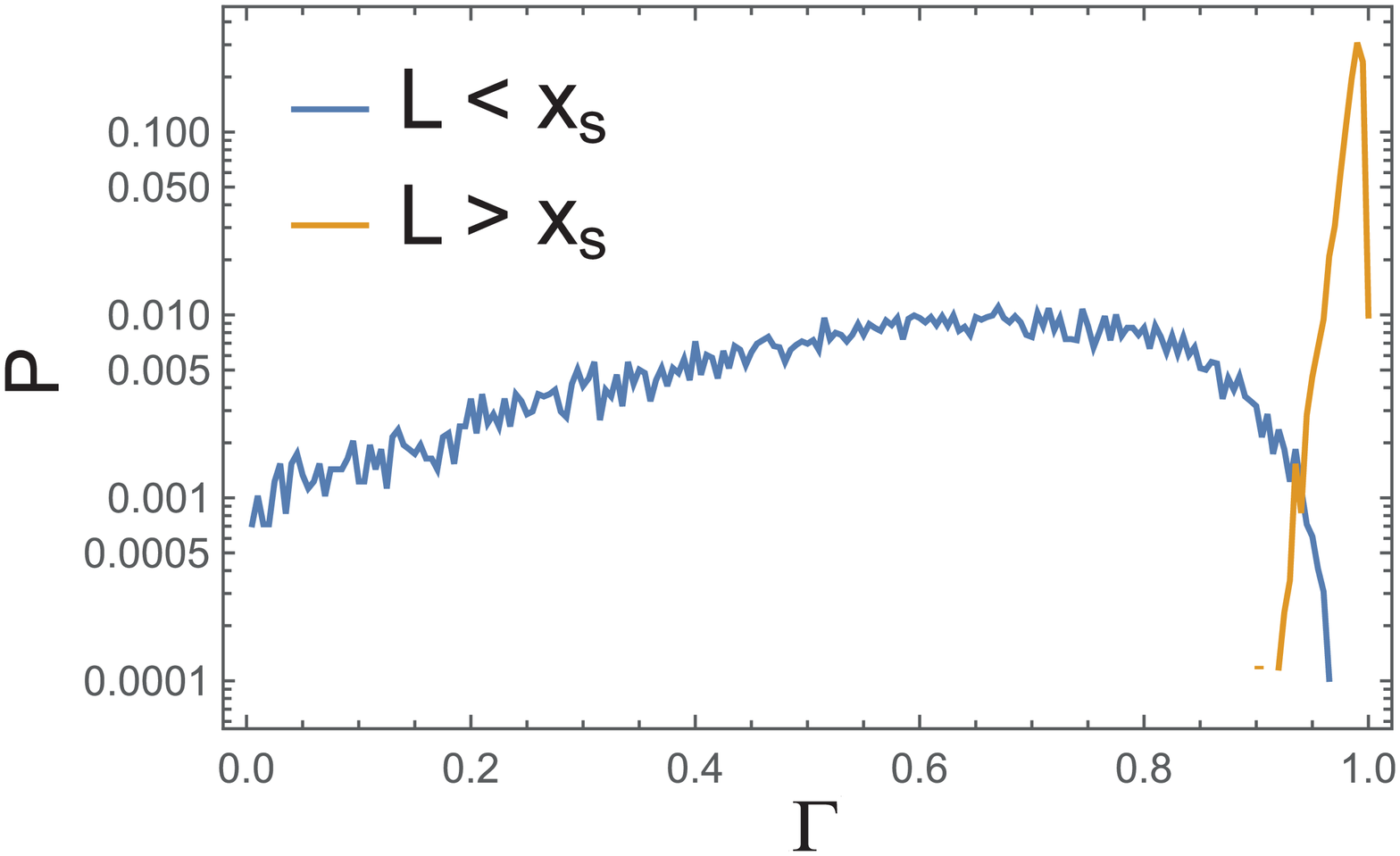,width=5cm}
\caption{Probability distribution $P(\Gamma)$ of a realisation-dependent correlation coefficient $\Gamma$, \eq{sc06}, for the trajectories starting outside and inside the correlated region $x_s$, \eq{sc05}. For the trajectories starting sufficiently far away from $x_s$, $P(\Gamma)$ is peaked at some value of $\Gamma$ which is close to $1$, implying that the dynamics of a tracer particle
is strongly influenced by the curved geometry of a channel. In this case, all the trajectories follow predominantly the same path. For $L < x_s$, a focusing of the trajectories does not take place and one observes a significant spread in values of the correlation coefficient.}
\label{fig03}
\end{figure}

Effective thickness $x_s$ of the layer within which the part $A'A''$ of the biased random walk is localised, can be estimated by extending the arguments presented in \cite{Nechaev2019} to determine conformations of a stretched polymer interacting with a convex impermeable wall. From an auxiliary geometrical construction presented in \fig{fig02}b we infer that
\begin{multline}
x_s\equiv |LM| = \sqrt{R^2 - |KM|^2} \\ = \sqrt{R^2 - (R-y_s)^2}\,\Big|_{y_s\ll R} \approx \sqrt{2 R y_s} \,,
\label{sc02}
\end{multline}
which relates the typical thicknesses of boundary layers in the $x$- and $y$-directions. In turn, time $t_s$ needed to move ballistically under the action of a reduced force $\mu$ on the distance $x_s$ obeys
\be
t_s = \frac{x_s}{\mu}
\label{sc03}
\ee
On the other hand, within this time interval the trajectory deviates in direction perpendicular to the boundary at a typical distance
\be
y_s=\sigma \sqrt{t_s}
\label{sc04}
\ee
Combining these estimates, we thus conclude that
\be
x_s \approx \sqrt[3]{\frac{4 \sigma^2}{\mu}} R^{2/3}; \quad y_s \approx \sqrt[3]{\frac{2 \sigma^4}{\mu^2}} R^{1/3}; \quad t_s \approx  \sqrt[3]{\frac{4 \sigma^2}{\mu^4}} R^{2/3}
\label{sc05}
\ee
Note next that the parameter $x_s$ determines the minimal distance to the target, in the longitudinal direction, within which the geometry of the system should be very essential. It is natural to expect that $x_s$ thus plays the role of a certain correlation length, and the behaviour of different realisations of trajectories within $x_s$ should be very similar. To quantify the degree of such a similarity, we study correlations between two different realisations of trajectories of the same length. Let ${\bf r}_i(N)$, $i=1,2$, denote two such $N$-step trajectories, which are completely specified by the positions of the intermediate points  $\{X_i,Y_i\}_{i=\overline{1,N}}$ in the plane $(x,y)$. Due to an external force pointing in the longitudinal direction, the $x$-coordinates of these two trajectories ${\bf r}_1(N)$ and ${\bf r}_2(N)$  have to be strongly correlated. For the $y$-component, the situation is less evident. Let us define the correlation coefficient of the $y$-components of two realisations of trajectories as follows:
\be
\Gamma=\frac{\sum\limits_{i=1}^N\left(Y^{(1)}_i-\langle Y^{(1)} \rangle\right)\sum \limits_{j=1}^N\left(Y^{(2)}_j-\langle Y^{(2)} \rangle\right)} {\sqrt{\sum\limits_{i=1}^N\left(Y^{(1)}_i-\langle Y^{(1)} \rangle\right)^2\sum\limits_{j=1}^N\left(Y^{(2)}_j-\langle Y^{(2)} \rangle\right)^2}}
\label{sc06}
\ee
where $Y^{(i)}_{i=1,2}$ are the $y$-coordinates of two paths ${\bf r}_1$ and ${\bf r}_2$. Note that
$\Gamma$ is a random variable which varies from a realisation to realisation. In \fig{fig03} we depict its distribution function $P(\Gamma)$ in two situations: for $L > x_s$ and $L < x_s$. We observe that in the latter case the distribution is rather broad which implies that values of $\Gamma$ are defocussed and the $y$-components of the trajectories are de-correlated. In contrast, when the distance between the starting point and the target exceeds the correlation length, the distribution is sharply peaked at some value of $\Gamma$ which is close to $1$. This signifies that different realisations of the tracer particle trajectories follow almost deterministically the same most probable path, i.e. are strongly correlated.

\begin{figure}[ht]
\epsfig{file=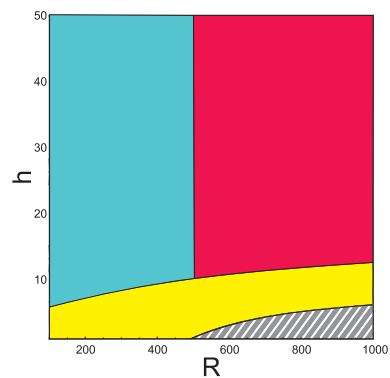,width=5cm}
\caption{Comparison of regions with a different behaviour of the tracer particle trajectories: a blue (upper left) region represents a domain in the parameter space in which a geometry-induced focusing takes place. A dashed domain shows the region in which the FJ approximation is valid.
Yellow and red areas indicate the domains in which a significant spread of the values of $\Gamma$ is observed.}
\label{fig04}
\end{figure}

The following question is quite legitimate: does the region of a geometry-induced focusing \eq{sc05} overlaps with a region in which the FJ approximation is valid \eq{fj04}? In \fig{fig04} we mark four different regions on the $(h,R)$-plane, for the trajectories starting at distance $L=100$ away from the bottleneck. Condition $\eta<1$ (recall that $\eta$ is the Fick-Jacobs parameter defined in \eq{fj04}), which is necessary for the FJ approximation to be justified, fulfils in the dashed grey region. In turn, a blue (an upper left) region in \fig{fig04} represents the domain in the parameter space in which a geometry-induced focusing is observed. We thus conclude that the effect of a focusing of the trajectories takes place away from the applicability region of the FJ approximation. Moreover, we argue that neither the original FJ approximation, nor its versions with $x$-dependent diffusion coefficient can match such a focusing effect.

\subsection{First-crossing heights for the translocation process}

The evolution of the tracer particle trajectories depends, in general, on many parameters: a distance between the starting point and the boundary, the magnitude of the external force, the channel's geometry, as well as the distance from a starting point to a funnel's centre. Here, we are mostly interested in the effects of a bottleneck thickness $H$ and of the location of the starting point on the dynamics. We thus restrict our analysis to these two parameters only.

\begin{widetext}

\begin{figure}[ht]
\epsfig{file=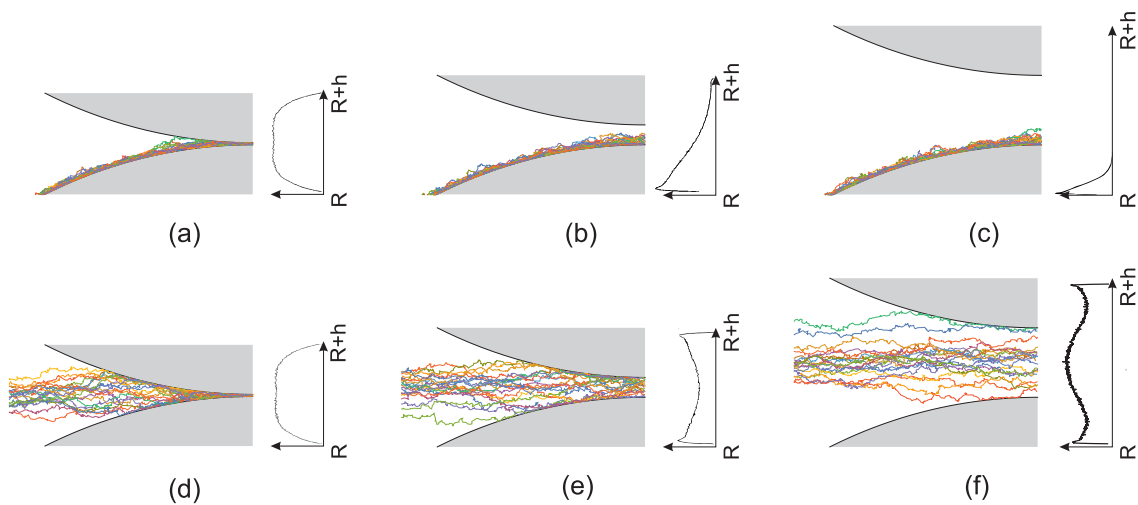,width=15cm}
\caption{Several typical realisations of trajectories of the tracer particle starting away from a funnel and passing through it. In the upper row we present trajectories that start at point $A$  (see \fig{fig01}),  while in the lower row - trajectories that start  on a principal axis of a channel (point $B$  in \fig{fig01}).  In all cases, the initial distance $L$ from the starting point to the center of the bottleneck is fixed $L=500$, the radius of a semicircle $R=500$, and the correlation length $y_s=10$. Panels (a), (b) and (c) (and respectively, (e), (f) and (d)) correspond to different thicknesses of the most narrow part of a bottleneck; $H=2$ for $(a,d)$, $H= 20$ for $(b,e)$ and $H=100$ for $(c,f)$. The graphs presented near each panel show the distribution $P(h)$ (obtained from a statistical ensemble of trajectories) of the height $h$ above the funnel's lower boundary at which a first crossing of a centre of a bottleneck occurs.}
\label{fig05}
\end{figure}

\end{widetext}

In \fig{fig05} we depict typical trajectories of a tracer particle starting away from the centre of a funnel and passing through the most narrow part of a bottleneck. In the upper row all the trajectories start at a point $A$ close to a boundary (see \fig{fig01}), while the lower row - they start right on the principal axis of a channel (point $B$ in \fig{fig01}). The distance $L$ from the starting point to the centre of a funnel is fixed and equal $L=500$, and the radius of obstacles -- semicircles -- in all the cases is the same, $R = 500$. The behaviour presented in panels (a), (b) and (c) (and respectively, (e), (f) and (d)) correspond to different thicknesses of the most narrow part of a bottleneck: $H=2$ for $(a,d)$, $H=20$ for $(b,e)$ and $H=100$ for $(c,f)$, and graphs near each panel show the distribution function $P(h)$ of the height above the lower boundary of a funnel at which a crossing of its most narrow part takes place.

From \fig{fig05} we can draw the following conclusions:

\noindent 1. For very narrow bottlenecks (panels (a) and (d)), when $H \ll y_s$, the trajectories starting from point $B$ move only along the boundary. The trajectories starting from the point $A$ reach the funnel's boundary sufficiently fast and predominantly keep on moving along it. The distributions $P(h)$ of the height $h$ at which a first crossing occurs are very similar for both starting points; $P(h)$ vanishes close to both boundaries and is nearly uniform (with some incipient deep in the middle) in the central part.

\noindent 2. For bottlenecks with some intermediate thickness, when $H \approx y_s$, the precise location of the starting point matters. Here, all the trajectories of a tracer particle which start at the boundary (panel (b)) are concentrated on the corresponding boundary, and the distribution $P(h)$ is strongly skewed -- it has a pronounced maximum on the boundary and rapidly vanishes with an increase of $h$. For the starting point located on the principal axis of a channel (panel (e)), the trajectories spread diffusively in the $y$-direction sufficiently fast to reach both boundaries and keep on moving along them. As a consequence, $P(h)$ is symmetric with respect to the principal axis of a channel and has a characteristic $M$-shaped form; it has two maxima at the boundaries and a deep in the centre.

\noindent 3. For wide bottlenecks, when $H \gg y_s$, the effect of the location of the starting point is most important. Here, the trajectories starting at the point A in \fig{fig01} run predominantly along this boundary (panel (c)); they do not go away to the bulk and thus
do not reach the opposite wall of the channel. As a consequence, the distribution of the height $h$ above the boundary at which a first-crossing of the centre of a funnel occurs is sharply peaked at this boundary. Note that in this case, the height fluctuations obey:
\be
\sqrt{\overline{h^2}-\overline{h}^2}\sim R^{1/3}\qquad \text{for $L > x_s$}
\label{sp01}
\ee
In contrast, the trajectories starting on the principal axis of a channel (panel (f)) are too short
to concentrate completely on the boundaries. Instead, they split into three populations -- two populations move along the boundaries, while one stays on the principal axis of a channel. The distribution $P(h)$ thus acquires a remarkable three-modal shape.

\begin{figure}[ht]
\epsfig{file=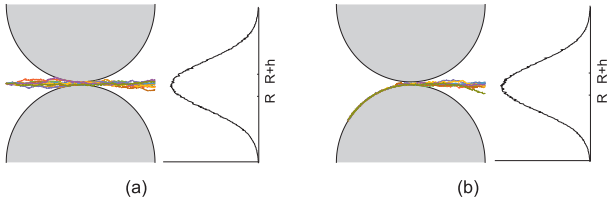,width=8cm}
\caption{Spread of the trajectories after a passage through a bottleneck. Panel (a): trajectories starting at point $B$. Panel (b): trajectories starting at point $A$. Graphs show the distribution of the height $h$ at which the trajectories appear at the end of a curved funnel -- at the entrance to a broad chamber.}
\label{fig06}
\end{figure}

Lastly, we address a question of the evolution of the tracer particle trajectories after a passage through a bottleneck. In \fig{fig06}, for the case when $H$ is of order of the correlation length $y_s$ (see \eq{sc05}), the show the spread of the trajectories at the entrance to a broad chamber.
We observe that $P(h)$ at this point acquires a bell-shaped form centred at the principal axis of a channel, such that all the asymmetry of this distribution emerging in the most narrow part of a bottleneck fades away. One expects, of course, that this asymmetric behaviour will be reproduced in the passage through the next bottleneck.

\subsection{First-crossing times for the translocation process}

Let $\tau$ be a random variable, defining the first-passage time to a target placed at the origin for a 1D Brownian motion, from a point $x_0 > 0$ and evolving in presence of a bias pointing towards the target. The variable $\tau$ has the following probability density function (see e.g.\cite{Mattos2012,MM2011}):
\be
P(\tau)=\left(\frac{b}{c}\right)^{\nu/2}\frac{1}{2K_{\nu}(2\sqrt{bc})}
\frac{1}{\tau^{1+\nu}}e^{-b/\tau-c \tau},
\label{fpt01}
\ee
which is called the generalised inverse Gaussian distribution. In \eq{fpt01}, $K_{\nu}(...)$ is the McDonald function, $\nu$ is the so-called persistence exponent \cite{SM1999} (for Brownian motion $\nu = 1/2$), $b=x_0^2/4D$, and $c=4D/\mu^2$, where $\mu$ is a reduced force acting on the tracer particle. As discussed in Refs. \cite{Mattos2012,MM2011}, the expression in \eq{fpt01} is quite  generic and we use it here for the interpretation of our numerical data. Using $b$, $c$ and $\mu$ as fitting parameters, we obtain their dependence on system's geometry, magnitude of the external force, starting position, etc.

\begin{figure}[ht]
\epsfig{file=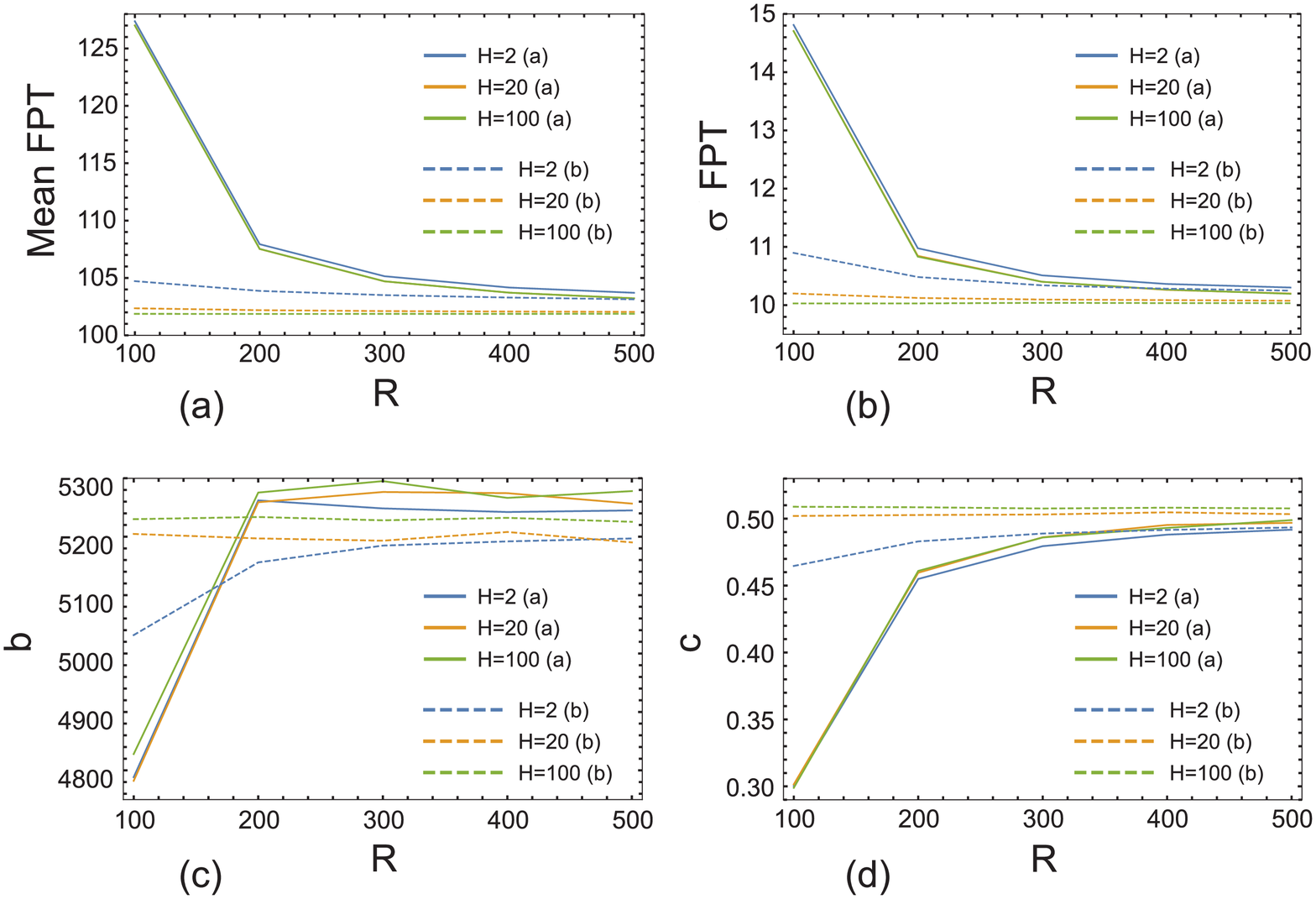,width=8cm}
\caption{Parameters of the probability distribution $P(\tau)$ in \eq{fpt01} as functions of $R$ for different values of a bottleneck thickness $H$. Solid curves correspond to the starting point located on the principal axis of a channel (point $A$), while the dashed ones -- to the starting point located on a funnel's boundary (point $B$). Panel (a): Mean first passage time; Panel (b): Standard deviation of the first passage time; Panels (c) and (d): parameters $b$ and $c$ of the
distribution function \eq{fpt01}.}
\label{fig07}
\end{figure}

In \fig{fig07} we summarise the results of numeric simulations which show a typical behaviour of the first passage times for two situations: (i) trajectories starting from the boundary (the point A in \fig{fig01}), when the length of the arc between the starting point and the target is equal to $L$, and (ii) trajectories starting on a principal axis of a channel at a distance $L$ away from the target (the point B in \fig{fig01}). The thickness of a bottleneck, $H$, is chosen according to the following prescriptions: $H=100$ corresponds to a channel in which a trajectory starting from a centre does not reach the channel's boundaries; $H=20$ corresponds to some intermediate thickness of a bottleneck -- it exceeds the correlation length $y_s$ but is smaller than a typical magnitude of fluctuations of trajectories in the tangential direction, and $H=2$ corresponds to a most narrow bottleneck, such that the tracer particle experiences multiple interactions with the boundaries before it crosses the most narrow point of a funnel.

We observe that in case when the starting point is located on the principal axis of a channel,
the mean passage time and the standard deviation in \fig{fig07} are decreasing function of $R$ (the radius of semicircular posts), and are independent of the thickness $H$ of a bottleneck. In contrast, for a starting point located on a boundary, these properties show a very small variation with $R$, and are apparently $H$-dependent. Further on, we realise that the distribution $P(\tau)$ is effectively \textit{narrow}. In standard statistical analysis (see e.g.\cite{Mattos2012,MM2011}) an effective broadness of some distribution function is characterised by the so-called coefficient of variation $\gamma_v$, which is defined as the ratio of the standard deviation and of the mean value. If the value of this parameter is less than unity, one deals with an effectively narrow distribution, while in case when it exceeds unity, fluctuations are important and the random variable cannot be characterised by its first moment only. We conclude from \fig{fig07} that for the two situations under study with fixed $L$ we have the coefficient of variation less than $1$. We hasten to remark, however, than even for the case at hand we cannot rule out observing higher values of $\gamma_v$, if we consider smaller values of $L$ (see Ref. \cite{mal4}, which studied such effects for an unbiased motion in a corrugated channel).

\begin{figure}[ht]
\epsfig{file=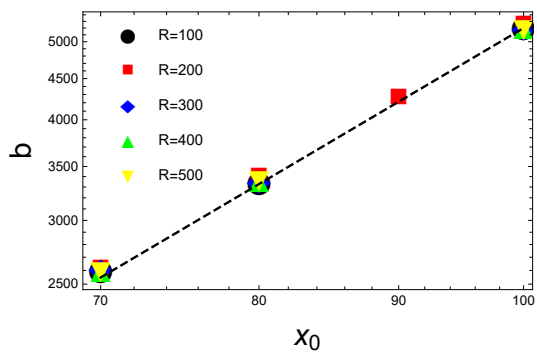,width=7cm}
\caption{Dependence of the parameter $b$ on the distance between the starting point and the centre of a bottleneck for different funnel radii. The dashed line is $x_0^2/2$.}
\label{fig08}
\end{figure}

Recall now that in one-dimensional systems, a parameter $b$ in \eq{fpt01} is proportional to a squared initial distance to the target. In a drift-dominated regime, trajectories are strongly stretched in the direction of a force, such that their longitudinal size is much larger than the transverse one. On an intuitive level, we thus can expect some similarity with a one-dimensional behaviour, regarding the dependence of $b$ on the distance $x_0$ to a starting point. To this end, in  \fig{fig08} we depict the dependence of $b$ on $x_0$ for four different values of $R$. We observe a perfect agreement with a one-dimensional behaviour $b = x_0^2/2$, regardless of the value of $R$.

\section{Conclusions}
\label{conc}

In this paper, we studied different characteristic features, both temporal and spatial, of a biased translocation process in a two-dimensional periodically-corrugated channel enclosed by hard-wall boundaries. The channel was modelled as a linear sequence of broad chambers connected sequentially by narrow bottlenecks -- funnels formed by two hard-wall semicircles (posts) separated from each other by some distance $H$ between their apex points. Having in mind typical settings of a constant force active micro-rheology, we focused on the dynamics of a tracer particle, biased by an external force to move along the channel, while undergoing an unbiased diffusive motion across the channel. This latter circumstance makes our model different from the analyses based on the Fick-Jacobs approximation, and our focal question here was what kind of information about crossings of the funnels is lost when one discards a possibility of the diffusion in the tangential direction.

More specifically, we analysed here, using scaling arguments and extensive numerical simulations, two distribution functions. The distribution function of the height above the boundary at which a first crossing of the most narrow part of a given funnel takes place, starting from a prescribed position, and the distribution of the time of a first crossing event (at any height).

Our analysis revealed a rather rich behaviour. We found that for bottlenecks of an intermediate thickness, comparable to a certain correlation length, a typical passage through a bottleneck occurs via sliding trajectories which travel very closely to either of the boundaries. For thicker bottlenecks, fine details of a first crossing event depend on the precise location of a starting point: if a particle starts in a vicinity of either of the boundaries, it travels along the boundary due to an emerging force-induced effective attraction to the  latter. In case when a particle starts on the principal axis of a channel, the trajectories split into three populations - two of them predominantly travel in a vicinity of the boundaries, while one fraction of trajectories continues to follow the principal axis. As a consequence, the resulting distribution of the height at which a first-crossing takes place has a spectacular three-modal form. Eventually, for very narrow channels, the distribution of a first-crossing height is defocussed independently of the location of a starting point -- this distribution is almost flat in its central part such that any value of the height in this region is equally probable.

Such a kind of passages through a given bottleneck is clearly a generic feature, and should be reproduced in crossings of all the subsequent bottlenecks. It may have important consequences: In particular, if channel's boundaries contain some chemically active species and a tracer particle may react with them, as it happens in some applications\cite{Schuss2007}, the reaction probability should be enhanced due to a sliding along the boundary. Moreover, because such channels in realistic systems are typically rough on a microscopic scale, particle travelling along a boundary will probe its microscopically rough landscape, which will increase the frictional force exerted on it, as compared to passages through a bulk filled with a solvent. As a consequence, for passages through a bottleneck the relevant parameter should be the surface mobility, rather than the mobility in the bulk.

Next, we analysed numerically the distribution of the first-crossing times. Fitting the obtained curves by the generalised inverse Gaussian distribution, we examined the dependence of its parameters on the parameters of our model.

We end up by noting that recently a very similar model has been considered\cite{active}, in which  the interior of a periodically corrugated channel was supposed to contain some amount of neutrally buoyant colloidal particles suspended in a solvent. In this insightful work the analysis was focused on the force-velocity relation -- the dependence of the terminal velocity of the tracer particle on a magnitude of the applied force. It is well-known that in such systems the interactions between a biased tracer particle and colloids\cite{benichou0,benichou, benichou1, benichou2} and also between biased tracer particles themselves\cite{rein,carlos1,carlos2, clog, kustner}, if several of them are present in the system, entail a spectacular and essentially cooperative behaviour. Such a behaviour takes place even in case of flat boundaries and originates
from emerging long-ranged, non-Newtonian interactions\cite{lowen}. In our future work we plan to examine the effects of curved boundaries on such out-of-equilibrium interactions along the lines of our present paper.

\begin{acknowledgements}
We wish to thank A. Gorsky, J.-F. Joanny, B. Meerson and K. Polovnikov for helpful discussions and valuable comments. S.N. and A.V. acknowledge the BASIS Foundation grant 19-1-1-48-1. A.V. and V.A. are supported by the state task for the FRC CP RAS \# FFZE-2019-0016.
\end{acknowledgements}

\bibliographystyle{plain}
\bibliography{channel-ref}

\end{document}